\documentclass{PoS}

\newcommand{\half}{\textstyle{ \frac{1}{2}}} 

\title{What is QFT? Resurgent trans-series, Lefschetz thimbles, and new exact saddles}

\ShortTitle{Resurgence and  Lefschetz thimbles}

\author{Gerald V. Dunne\\
          \\Department of Physics, University of Connecticut, Storrs, 
CT 06269-3046, USA \\
        E-mail: \email{dunne@phys.uconn.edu    }}
        
\author{\speaker{Mithat \"Unsal}
 \\
Department of Physics, North Carolina State University, Raleigh, NC 27695, USA \\
        Department of Mathematics, Harvard University, Cambridge, MA, 02138, USA
         \\
        E-mail: \email{unsal.mithat@gmail.com}}

\abstract{This is an introductory level review of recent  applications of resurgent trans-series and Picard-Lefschetz theory to quantum mechanics and quantum field theory.   Resurgence connects local perturbative data with global topological structure.  In quantum mechanical systems, this program provides a constructive relation between different saddles. For example, in certain cases it has been shown that all information around the instanton saddle is encoded in perturbation theory around the perturbative saddle.  In quantum field theory, such as sigma models compactified on a circle,  neutral bions  provide a semi-classical interpretation of the elusive IR-renormalon, 
and fractional kink instantons lead to the non-perturbatively induced  gap, of order of the strong scale.  In the path integral formulation of quantum mechanics, saddles must be found by solving the holomorphic Newton's equation in the inverted (holomorphized) potential. Some saddles are complex, multi-valued, and even singular, but of finite action, and their inclusion is strictly necessary to prevent inconsistencies. 
The multi-valued saddles enter either via resurgent cancellations, or their phase is tied with a hidden topological angle.  We emphasize the importance of the  destructive/constructive interference effects between equally dominant saddles in the Lefschetz thimble decomposition.  This is especially important in the context of the sign problem. 
 }

\FullConference{The 33rd International Symposium on Lattice Field Theory\\
		14 -18 July 2015\\
		Kobe International Conference Center, Kobe, Japan*}

\begin{document}
\begin{quote}
``Nearly all   natural problems with analytic data but without analytic solutions lead to formal power series which turn out to be resurgent in some appropriate variable. Viewed from that angle,  the world of  resurgent functions is almost coextensive with the world of divergent expansions." 
\end{quote} 
\vspace{-0.3cm}
 \hspace{6 cm}    J. Ecalle   {\em {Les Fonctions Resurgentes, Vol.-III}} \\

\vspace{-0.6cm}
\section{Introduction}
In this brief review,  we provide   an introduction to the basic  ideas underlying  the  application of  resurgence,  trans-series  and Lefschetz thimbles to  problems in  quantum mechanics (QM), and  quantum field theory (QFT).   These tools have the potential to provide a new non-perturbative definition of QFT in the continuum, and which may also be computationally practical. A related goal is to find new ways to perform numerical simulations of path integrals, for example for path integrals  with a sign problem, such as finite density QCD. These methods also provide a new perspective on semi-classical methods in path integration, and lead us to reconsider from scratch the basic textbook approach to semi-classics.
Our goals are not specific to any particular QFT;  rather we aim to find techniques of universal character, providing insights to all QFTs. 

\vspace{3mm}

The ideas from mathematics come from two  beautiful and powerful  
notions:
\begin{itemize}
 \item  Resurgent  trans-series  which go beyond  conventional 
 asymptotic analysis \cite{Ecalle-book, Berry657}.
 \item   Picard-Lefschetz theory,  a complex version of Morse theory 
  \cite{Fedoryuk}. 
 \end{itemize}
 These two methods,  one related to local analysis,  and the other related to global topology, despite sounding extremely  different,  are actually deeply related.\footnote{Resurgent analysis may possibly be even more general \cite{Ecalle-book}.  There are many problems which do not seem to arise from integration,  or do not admit a saddle point expansion, at least in any obvious way.  Yet, there is still evidence that a resurgent structure exists.}  At least for finite dimensional exponential type  integrals,  
$\int dx_1 \ldots dx_N  \; e^{-\frac{1}{\lambda} f(x_1, \ldots, x_N)}$, where $f(x_1, \ldots, x_N)$ is meromorphic,  local analysis and global topology seem to be two  sides of the same coin.  In infinite dimensional path integrals, the extent of the  relation  between the two is an open problem, which attracts the attention of mathematicians and physicists alike  \cite{Kontsevich-1,Argyres:2012ka, Dunne:2012zk,Dunne:2013ada}.
Our  work in this direction may be viewed in two different ways. One is extracting new physical phenomena that can be found from  a  proper semi-classical analysis of the path integral, and   the other is what can be viewed as  "experimental mathematics", i.e., finding new structure of the path integral and communicating this understanding to mathematicians.   We  hope that this will lead to both mathematical and physical insights into the problem of path integration.  Specifically, we anticipate new ideas concerning:
\begin{itemize}
\item
the practical geometric structure of field space (the surprise in resurgence is that it connects  local analysis to global structure/information),
\item
systems with an infinite number degree of freedom.
\end{itemize}
  
  \vspace{3mm}

We discuss here insights from physics coming from two main sources: 
 \begin{itemize} 
  \item Exact/uniform WKB, or exact quantization conditions in QM, and their path integral implications \cite{Dunne:2013ada}.
 \item Semi-classically calculable compactifications and deformations of  QFTs, systems with infinitely many degrees of freedom  \cite{Dunne:2012zk, Unsal:2007jx, Unsal:2008ch}.
\end{itemize}
Calculable deformations  and compactifications provide  new windows into gauge theories
on $\mathbb R^3 \times S^1$   \cite{Unsal:2007jx,  Argyres:2012ka, Bhoonah:2014gpa},
 and into sigma models on $\mathbb R^1 \times S^1$ \cite{Dunne:2012zk,Cherman:2013yfa}.   This step defines a regime in QFT in which semi-classical analysis is reliable,  and the theory is continuously connected to the strongly coupled regime on   $\mathbb R^4$ and $\mathbb R^2$, respectively. This  gives a new semi-classical interpretation of the elusive infrared renormalon of 't Hooft, in a weakly coupled regime.  This formalism also reveals  a wide new class of non-perturbative saddles, such as charged/magnetic and neutral bions, and their roles in dynamics, e.g. mass gap generation,  center symmetry  and  chiral symmetry realization.

Exact WKB/exact quantization conditions  provide guidance  to  why observables in QM should  be resurgent \cite{ddp,ZinnJustin:2004ib, Aoki:1993ra}.  A new application of uniform exact WKB yields 
a constructive formula that means that one can deduce all non-perturbative data from perturbation theory, in certain QM cases  \cite{Dunne:2013ada}. This is quite different from other forms of resurgence, 
which typically connect low orders of perturbation theory around instanton saddles, to low orders of perturbation theory around the perturbative saddle.
Uniform WKB also has a natural geometric aspect; in particular, it directly yields  the  correlated multi-instanton event amplitudes  \cite{ Dunne:2013ada, Misumi:2015dua}.  In general, 
   calculating the multi-instanton amplitude used to be done using the Bogomolny/Zinn-Justin (BZJ) prescription  \cite{Bogomolny:1980ur}, but this is not always satisfactory,  and can even lead to wrong results if not interpreted carefully.   The result from uniform WKB turns out to be  consistent with multi-instanton calculus if 
   one uses a  finite dimensional version of  Picard-Lefschetz theory  for the Lefschetz thimble description 
of the quasi-zero mode integration cycles \cite{Behtash:2015kna}. Furthermore, doing the quasi-zero mode integration via  Lefschetz thimbles fixes the problems with the  BZJ-prescription. 

Additional motivation and intuition comes from the many interesting results in recent applications of resurgence to matrix models, topological string theory and localizable SUSY gauge theories \cite{Marino:2007te,Marino:2012zq,Aniceto:2014hoa,Aniceto:2013fka}.

\section{Saddle point method, and geometrization of Borel resummation}
The saddle point method is standard textbook material, but it is actually much deeper than is presented in many books, which generally only consider local quadratic fluctuations.
The method has two different parts: one is tied with local analysis, and the other with topology. Asymptotic analysis around a given critical point  of the action concerns a perturbative expansion around the given point. The topological part concerns the deformation of the integration cycles, and also the homology cycle decomposition of the original integration cycle. Our description of both steps will be conceptual; see \cite{Berry657,Fedoryuk,Kontsevich-1} for details. 

Consider a finite-dimensional exponential integral of the form 
\begin{eqnarray}
\int_{\Sigma} dx_1 \ldots dx_N  \; e^{-\frac{1}{\lambda} S(x_1, \ldots, x_N)} \qquad 
\end{eqnarray}
 where  $\lambda$ is  a small parameter, and $\Sigma$ is an integration cycle to be determined. 
 Even if an ordinary integral is formulated over real variables, the natural space in which 
the critical points (saddles) $\rho_\sigma$, and their critical point cycles  ${\cal J}_\sigma$,
 live  is the {\it complexification}  of the original space.  Complexification doubles the dimension,  but the critical point cycles ${\cal J}_\sigma$ are middle-dimensional: their dimension is the same as  the original space, or half that of the complexified "field space".  

As an  example, for an ordinary integral over  $N$-dimensional real space, this procedure is 
\begin{eqnarray} 
\Sigma^N \subset \mathbb R^N 
\longrightarrow   X  = \mathbb C^N \longrightarrow  \Sigma^N  = \sum_\sigma n_\sigma {\cal J}_\sigma,
\end{eqnarray}
 where the  original integration cycle $\Sigma^N$ may be viewed as a real locus on the complex algebraic variety $X= \mathbb C^N$: 
 \begin{eqnarray}
  \Sigma^N 
= \sum_\sigma n_\sigma {\cal J}_\sigma, \qquad   \dim_{ \mathbb R} ({\cal J}_\sigma)
= N.
\end{eqnarray}
For $N=1$ this simply means that the original 1-dimensional integral cycle becomes a sum over deformed 1-dimensional contours in the (2 real dimensional) complex plane.
For finite dimensions greater than 1, this decomposition becomes a very interesting problem, and the determination of the thimbles amounts to solving the   complex gradient flow equation \cite{Fedoryuk}. The  infinite dimensional version of this construction was  introduced by Witten  for  phase space QM and QFTs  \cite{Witten:2010zr} (see also \cite{Guralnik:2007rx,Tanizaki:2014xba}).
For an ordinary integral, the above prescription amounts to 
\begin{eqnarray}
I(\lambda)=\int_{\mathbb R^N}  dx_1 \ldots dx_N \;  e^{-\frac{1}{\lambda}\, S(x_i)}  \rightarrow \sum_\sigma n_\sigma   \int_ 
{ {\cal J}_\sigma} dz_1  \ldots dz_N \;  e^{-\frac{1}{\lambda}\,S(z_i)}
\label{complex}
\end{eqnarray}
where ${ {\cal J}_\sigma}$ can be found by solving  a complex version of the gradient flow equations (also called the Picard-Lefschetz equation)
\begin{eqnarray}
\frac{\partial z_i }{\partial \tau} = 
      -  \frac{ \partial \bar S }{ \partial \bar z_i}\, , \qquad 
\frac{\partial  \bar z_i }{\partial \tau} = 
      -  \frac{ \partial S }{ \partial z_i}  \, , \qquad (i=1, \dots, N), 
\label{PLW}
\end{eqnarray}
where $\tau$ is the flow time, and the over-bar means complex conjugation. Using (\ref{PLW}) and the  chain rule, 
$
\frac{\partial  {\rm Im} [S] }{\partial \tau} =   
\frac{1}{2i} \left( \frac{\partial S } {\partial z_i} 
     \frac{\partial z_i }{\partial \tau} 
 -   \frac{ \partial \bar S }{ \partial \bar z_i}   
     \frac{\partial\bar z_i }{\partial \tau}    \right) = 0
$: the imaginary part of the action is {\it invariant} 
under the flow
\begin{eqnarray}
{\rm Im} [S(z_1, \ldots, z_n)] ={\rm Im} [S_\sigma] 
\label{s-phase}
\end{eqnarray}
and (\ref{complex}) may therefore be re-written as 
\begin{eqnarray}
I(\lambda) =\sum_\sigma n_\sigma  e^{-\frac{i}{\lambda}  \; {\rm Im} [ S_{\sigma}] } \int_ 
{ {\cal J}_\sigma} dz_1  \ldots dz_N \;  e^{-\frac{1}{\lambda}\, {\rm Re}S(z_i)}   \equiv \sum_\sigma n_\sigma\,   e^{- \frac{i}{\lambda} \;  {\rm Im} [ S_{\sigma}] } I_\sigma (\lambda)
\label{complex-2}
\end{eqnarray}
where
$I_\sigma (\lambda)$ is the  finite result of the integration over the thimble  ${\cal J}_\sigma$.   

At this point,  we may  establish a connection between the Lefschetz thimble decomposition and resurgent trans-series. 
Locally, around a critical point,  $I_\sigma (\lambda)$ can be expanded into an asymptotic power series, given by  
\begin{eqnarray}
I_\sigma (\lambda)   \sim   e^{-\frac{1}{\lambda}\, {\rm Re} [ S_{\sigma}] }  \sum_p a_p^\sigma  \lambda^p= e^{-\frac{1}{\lambda}\, {\rm Re} [ S_{\sigma}] }  \Phi_\sigma (\lambda)
\end{eqnarray}
$ \Phi_\sigma (\lambda)$ is a formal asymptotic series, interpreted as the fluctuations around the saddle $\rho_\sigma$. The integration over the thimble ${\cal J}_\sigma (\theta)$ is equivalent to performing a directional Borel resummation of the formal power series in some direction $\theta$ in the Borel plane \cite{Ecalle-book}:
\begin{eqnarray}
I_\sigma (\lambda) =   \underbrace{  \int_ 
{ {\cal J}_\sigma (\theta) }  dz_1  \ldots dz_n \;  e^{-\frac{1}{\lambda} S(z_i)} }_{ \rm {integration \;  over \;  thimble}} =  \underbrace{e^{- \frac{1}{\lambda}\, S_{\sigma} }  { \cal S_{\theta} } \Phi_\sigma (\lambda)}_{\rm {Borel  \; resummation }} \qquad , \quad  \lambda \equiv |\lambda| e^{i \theta} 
\end{eqnarray}
In other words, the Borel resummation is equal to the integration over the thimble.  An intuitive reason why this is possible can be found in \cite{Berry657}, but ultimately ties up with the fact that the Stokes phenomenon is a co-dimension one effect. 

 Note that  the (directional) Borel resummation is  well-defined (unambiguous)  for all directions in which there is no singularity. In singular directions, Borel resummation is ambiguous: there is a Stokes jump between the left and right resummation, along directions just to the left or right of the singular direction. 
Analogously, the integration cycle is also ambiguous, and as one crosses a Stokes ray, the thimble exhibits a jump, 
\begin{eqnarray}
{\cal J}_{\sigma} (\theta_0^{-}) \rightarrow {\cal J}_{\sigma} 
 (\theta_0^{+}) + n_{\sigma \sigma'}{\cal J}_{\sigma'} (\theta_0^{+})  . 
\end{eqnarray}
For $N=1$, this formalism is the well-known stationary phase approximation, with the natural incorporation of  the Stokes phenomenon \cite{Ecalle-book, Berry657}. 
The  definition of the Lefschetz thimble based on stationary phase,  
${\rm Im} [S(z_i) -S_\sigma]=0$, is {\it only} satisfactory for a one-dimensional integral. In that case, 
this condition  provides one real condition on a one-complex dimensional space, and determines the thimble uniquely. 

For $N >1 $,    the integrand lives in a $2N$ (real) dimensional space, while the surface is $N$ real dimensional. The single stationary phase condition defines a 
co-dimension one (real dimension $2N-1$) space, which is not the desired $N$ real dimensional integration space. Instead, one 
needs $N$ real conditions to define the thimble. For finite dimensional integrals, Fedoryuk  defines
 a  steepest surface (descent manifold)  ${\cal J}_\sigma$   by  using the complex gradient system, 
  with initial conditions not lying at critical points (otherwise, no flow will occur) \cite{Fedoryuk}.

This means that unlike the $N=1$ case, for $N\geq 2$ it is not so easy to determine which critical points are relevant for the original integration cycle, $\Sigma^N$.  Howls proposes an algebraic approach to this  \cite{Berry657}, based on a hyperasymptotic expansion of the late terms. Remarkably, ideas very close to those of  \cite{Berry657} are shown to work with infinite dimensional  path integrals \cite{Basar:2013eka}. 
The analysis of  \cite{Basar:2013eka} shows that for a real periodic  potential (a Jacobi elliptic function) the large-order behavior of the perturbative expansion receives contributions from both real and complex saddles, even though the theory is completely real, and the underlying path integral is a sum over just real paths. Somehow perturbation theory ``knows'' that the real potential is also periodic in the complex direction, and so has complex instantons.
In other words, even though the Stokes multipliers of certain saddles are zero, their imprint in large-order perturbation theory is clearly present.

One other  difference between the $N=1$ and $N \geq 2$   case takes place once a Stokes phenomenon occurs. In $N=1$, when this happens, two critical points 
$\rho_\sigma$ and $\rho_{\sigma'}$  must lie on the same thimble.  In higher dimensions, two critical points no longer have to be on a hyper-surface which  passes through both 
$\rho_\sigma$ and $\rho_{\sigma'}$ simultaneously.   The  Stokes phenomenon in the $N \geq 2$   cases may just correspond  to the meeting of two separate thimbles. This meeting will generically happen at a  a lower dimensional hyper-surface (like the join of two fingers of a glove,  where the common vanishing cycle will no longer be smoothly defined).

Stokes phenomenon, even in multiple dimensional integrals, is a co-dimensional effect.  Only one free parameter (e.g. phase of the coupling) is sufficient to observe the phenomenon.  In this sense, 
the Borel plane structure is a {\it stratification} of saddles and their Lefschetz thimbles. For a nice discussion of this material, we refer to Ref.~\cite{Berry657}.

To summarize, the Lefschetz thimble or resurgent trans-series decomposition of an ordinary multi-dimensional exponential integral can in general be expressed as 
\begin{eqnarray}
I (\lambda) =  \sum_{\sigma \in  {\rm Active}}  n_\sigma    \; e^{-  \frac{i}{\lambda}\;  {\rm Im} [ S_{\sigma}] }  e^{- \frac{1}{\lambda}\, {\rm Re} [S_{\sigma}] }  { \cal S_{\theta} } \Phi_\sigma (\lambda)
\label{complex-2}
\end{eqnarray}
where $n_\sigma$ are piece-wise constant multipliers.  We sum over only active  (contributing) saddles in a given Stokes wedge, along with their multipliers. As one moves from one Stokes wedge to another, these coefficients will jump.  

The expansion (\ref{complex-2}) may be composed of hierarchical exponential factors,  and one may be tempted to think that just picking up the dominant saddle will suffice for asymptotic analysis. 
 But in a large-number of very interesting applications,  both in QFT, QM,  ordinary integrals as well as matrix models, there are equally dominant saddles (whose real part coincide) but whose phases differ. 
 In such cases, there will be extremely important destructive/constructive interference effects among saddles. The hidden topological angles in QM and QFT are examples of this type that we discuss in Section.\ref{HTA}.

 \begin{figure}[h]
     \includegraphics[width=1.\textwidth]{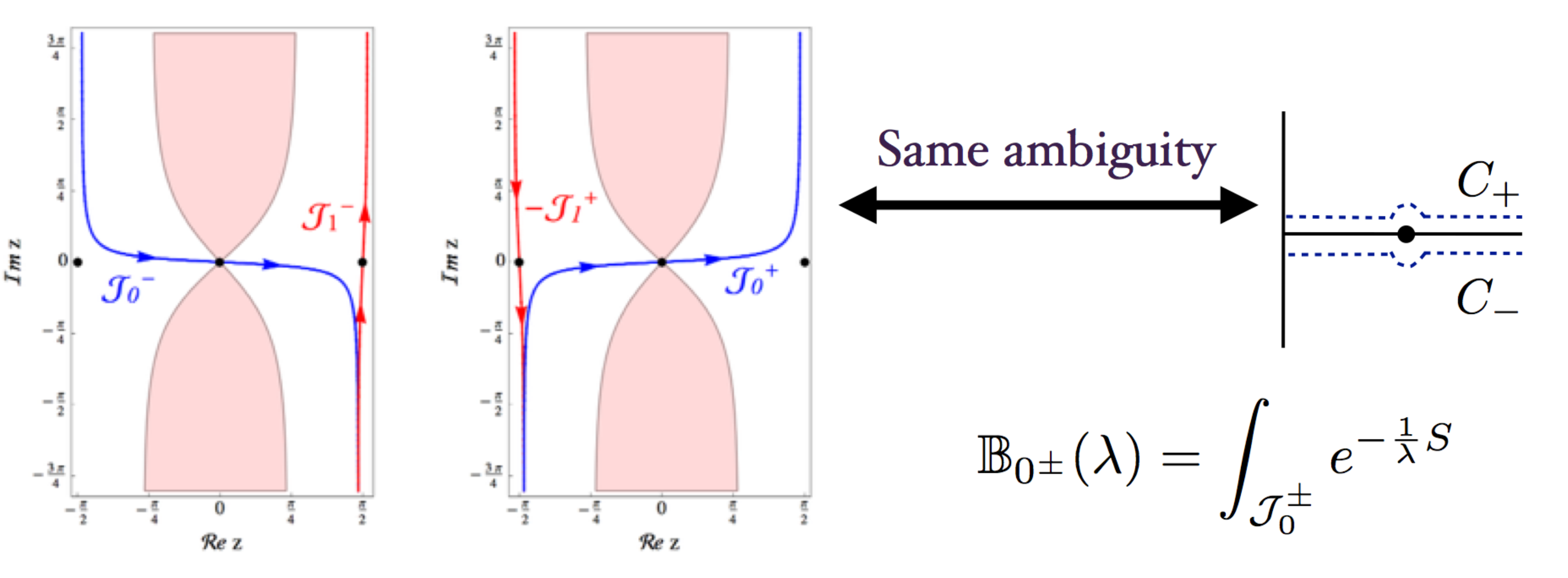}
     \caption{(Left:)  Critical points and their Lefschetz thimbles (descent cycles).  (Right:) Borel plane structure. Circle denotes a branch point.   Integration over thimbles is in one-to-one correspondence with directional Borel resummation.}
     \label{fig1}
     \end{figure} 
     
\subsection{Prototype ``$d=0$ path integral'' example:}
Consider a  ``zero dimensional path integral'',  the ordinary integral $Z(\lambda)= \frac{1}{\sqrt \lambda}  \int_{- \pi/2}^{ \pi/2} dx \; e^{- \frac{1}{2 \lambda} \sin^2 x}$ in the steepest descent method. There are two saddles: $z_0=0$ (perturbative), and $z_1= \pi/2$ (non-perturbative). To each saddle we associate 
a Lefschetz thimble (steepest descent path),  ${\cal J}_0 (\theta)$ and  and ${\cal J}_1 (\theta)$, respectively,  introducing $\theta=\arg (\lambda)$ to study properties of the integral under analytic continuation. This integral can be studied either by using perturbative expansions around the saddles, and subsequent Borel resummation of the associated asymptotic series, or by just performing the integrals over the steepest descent cycles.  These two procedures are actually the same. 

 By inspecting  Fig. \ref{fig1}, we observe that  the  original integration cycle $[-\pi/2, \pi/2) $ can be expressed at $\theta=0^{-}$ as  ${\cal J}_0 (0^{-})  + 
{\cal J}_1 (0^{-}) $. Upon  Stokes  jump,   the monodromy of the cycles crossing the Stokes ray 
$\theta=0$ are:
\begin{equation}
\begin{array}{l}
 {\cal J}_0    \longrightarrow   {\cal J}_0  -2  {\cal J}_1    \\
  {\cal J}_1    \longrightarrow {\cal J}_1   
\end{array} 
\qquad 
{\rm or}  \;\;  {\cal J}_i \rightarrow U_{ij}  (\theta=0) {\cal J}_j \qquad 
{\rm with }  \;\; 
 U_{\bf \circlearrowleft}(0)= \left(
\begin{array}{cc}
1  & -2   \\
  0&  1    
\end{array}
\right) \; 
\end{equation}
where $U_{\bf \circlearrowleft}(0)$  is  an upper triangular matrix. The  integration cycle at $\theta=0^{+}$ is  ${\cal J}_0 (0^{+})   - {\cal J}_1 (0^{+}) $.

  Note that the coefficient of the dominant  perturbative saddle does not jump, but the ${\cal J}_0 (\theta)$ cycle jumps at $\theta=0$, and in contrast,  the coefficient of the recessive non-perturbative saddle does jump, but the ${\cal J}_1 (\theta)$ cycle does not jump at $\theta=0$.
 
It is straightforward to show that $\int_{{\cal J}_0 (0^{+})} - \int_{{\cal J}_0 (0^{-})}$, i.e, the discotinuity in the 
integral over the two cycle is equal to  the difference of the left/right Laplace integral of the Borel transform $\int_{C{+}} -   \int_{C{-}} $, see Fig.\ref{fig1}. These are the left and right Borel resummations.  Indeed, by a simple change of variables, one can see that the integration over the thimble is actually 
Borel resummation. In this sense, the Borel plane pictures are a {\it stratification} of saddles and their Lefschetz thimbles. 
This idea for the multi-dimensional integrals  is expressed in Pham and Howls in more detail \cite{Berry657,Fedoryuk}. The reason this is doable in a multi-dimensional integral is because the 
Stokes phenomenon is a co-dimensional effect.  Only one free parameter (e.g. phase of the coupling) is sufficient to observe the phenomenon.  An analogous example with three saddles points is analyzed in detail in \cite{Basar:2013eka}.

\section{Resurgence triangle in quantum mechanics and two different types of resurgence}
In QM, there are two different types of resurgent relations.  In order to describe this, it is useful to define  the "graded resurgence triangle"  of QFT  \cite{Dunne:2012zk}, here for quantum mechanics \cite{Dunne:2013ada}. The resurgence triangle is a trans-series 
decomposition of the path integral, expressed in terms of cells  $(n, m), |m| \leq n$.
The rows are sectors with fixed action $(nS_I)$, $n=0, 1,2 ,\ldots $,  in units of the instanton action $S_I$, and the columns  are  sectors with fixed  topological charge $|m|  \leq n$.  The usual topological discussion in QFT and QM only distinguishes saddles according to topological charge, and all events in a given column are in the same topological sector. But resurgence, or a general saddle point decomposition,  provides a more refined construction. 

 The resurgence triangle for  the periodic potential (which also admits a topological charge) is given by 
\begin{eqnarray}
&1 f_{(0,0)} &\nonumber\\ 
\cr
  e^{-\frac{S_I}{g}+i  \theta } f_{(1, 1)}  \hskip -20pt
&& \hskip -20pt  \quad
e^{-\frac{S_I}{g} -i \theta } f_{(1, -1)} \nonumber\\ \cr
e^{-\frac{2S_I}{g}+2i \theta  } f_{(2, 2)}
\quad \quad & 
e^{-\frac{2S_I}{g}} f_{(2, 0)}  &
\quad \quad
e^{-\frac{2S_I}{g}-2i  \theta} f_{(2, -2)}
\nonumber\\  
 \cr
e^{-\frac{3S_I}{g}+3i   \theta} f_{(3,3)}
\quad \quad
e^{-\frac{3S_I}{g}+i   \theta  } f_{(3,1)}
 \hskip -20pt
&& 
e^{-\frac{3S_I}{g}-i  \theta} f_{(3,-1)}
\quad \quad
e^{-\frac{3S_I}{g}-3i   \theta} f_{(3, -3)} 
\label{triangle}
\end{eqnarray}
In the periodic potential case, we can introduce a topological $\theta$-angle. Since perturbation theory around any sector  is  independent of the $\theta$ angle, sectors with different $\theta$ dependence cannot mix with each other in the cancellation of the non-perturbative ambiguities.  
There are two essential type of non-perturbative ambiguities. 
 \begin{itemize}
 \item{Ambiguity associated with non-Borel summability of perturbation theory around the perturbative vacuum, and around any other saddle. Thus, for the QM periodic potential problem, all formal series $f_{(n,m)}$ are non-Borel summable, i.e, the sum is ambiguous.} 
 \item{Ambiguity in the definition of "$n$-defect"  amplitudes.  An n-defect is a correlated event of $n$ instantons and anti-instantons. The amplitudes at the edges of the triangle ($n$-strict instantons or anti-instantons) are ambiguity free (as can be deduced by integrating over the quasi-zero mode.) Indeed, this must be the case,  because these are the lowest action  configuration in the corresponding homotopy class. 
  All amplitudes in the interior of the triangle are ambiguous, but  in a calculable way.   }
 \end{itemize} 
 The information in the $m$-th column of the resurgence triangle is contained  in the twisted partition function 
 $Z_{m}$, with the insertion of the translation operator $T$.
 \begin{eqnarray}
Z(\beta, g, \theta)= \sum_{m=-\infty}^{+ \infty} e^{i m \theta} Z_{m} (\beta, g), \qquad 
 Z_{m} ={\rm tr} \; T^{ m} e^{-\beta H}=   \int_{x(t+ \beta)=  x(t) +  m \frac{\pi}{g} } Dx(t) e^{-\frac{1}{g}\, S[x]}  
\end{eqnarray}
{\bf Resurgence (conventional):}
Usual resurgence connects saddles within a given topological sector. 
For the cosine potential the late terms around the perturbative saddle $[0,0]$ diverge as \cite{Dunne:2013ada}
\begin{eqnarray}
c_n^{[0,0]} \sim \frac{n!}{{\color{blue}(2S_I)^n}}  \left( {\color{red}1} - {\color{red} \frac{5}{2} }\cdot \frac{{\color{blue}(2S_I)^1}}{n} - {\color{red} \frac{13}{8} } \cdot  \frac{{\color{blue}(2S_I)^2}}{n(n-1)} + \ldots \right)   
 \end{eqnarray}
 while the imaginary ambiguous part of the $[I\bar I]$  contribution to the ground state energy is
 \begin{eqnarray}
{\rm Im} [I\bar I]_{\pm} \sim  \pm \pi e^{ -{\color{blue}(2S_I) }/{g} }  \left( {\color{red}1}  - {\color{red} \frac{5}{2} }\cdot   g  - {\color{red} \frac{13}{8} } \cdot  g^2  \ldots \right)   
 \end{eqnarray}
 Note the correspondence between the coefficients in these two very different expansions.
This is a late term/early term relation in the following sense:
Late terms (around the dominant saddle) and the subleading $1/n$ corrections to them, encode information about  early terms in the perturbative, expansion in $g$ around a  subdominant saddle. Thus the $n!$-diverging late terms are not {\it meaningless or a nuisance} (as Berry puts it), but contain exact coded information about another saddle in the problem, in this case, the instanton-anti-instanton saddle. Similarly, the large-order behavior of the fluctuations about the single-instanton saddle are directly related to the low orders of the fluctuations about the $[  I I\bar I]$ saddle, and so on.
This structure is an imprint of resurgence, and for resurgent functions such relations exist essentially universally, intrinsically tied to the cancellations that define the trans-series as describing a unique well-defined function \cite{Aniceto:2013fka}. 

This result tells us that perturbation theory by itself is pathological, and so is the semi-classical 
 $[I\bar I]_{\pm}$ saddle. But in combination, the pathological parts ``cure'' each other. Namely, the imaginary ambiguous part associated with the Borel resummation of the perturbation theory, ${\mathbb B}f_{(0,0),\pm}$,  cancels against the ambiguity of the  $[I\bar I]_{\pm}$ amplitude:
 \begin{eqnarray}
{\rm Im} {\mathbb B}f_{(0,0),\pm} +  {\rm Im} [I\bar I]_{\pm}  {\mathbb B} f_{(2,0),\pm}   + \ldots   =0 
\label{cancel}
 \end{eqnarray}
 As a result, the topological charge zero column in the resurgence triangle gives a meaningful, real, and ambiguity free result. This is the sense in which resurgence renders the combination of the perturbative and non-perturbative expansion meaningful, and it may potentially provide a non-perturbative definition of path integral. 

This structure is extemely elegant; but  the story is  even more interesting as discussed next.
 
\vspace{0.3cm} 
\noindent 
{\bf Resurgence (new):}  
The conventional resurgent cancellations do not imply any relation between the perturbative vacuum saddle and the instanton saddle, or the two-instanton $[I I]$ saddle etc; in other words between different columns of the resurgence triangle (\ref{triangle}). In fact, there cannot exist a usual resurgence type relation between a perturbative saddle and an instanton saddle, because  perturbation theory is independent of the topological  theta angle while the instanton saddle depends on it. 
Thus a pathology of perturbation theory cannot be cancelled by a pathology of  the instanton saddle.    
Nevertheless,  there is such relation, it is simple and it is constructive. By constructive, we mean, for example, that if you know a certain number of orders of perturbation theory  around the perturbative saddle, you can immediately write down the same number of orders of the perturbation theory around the instanton saddle.  Namely, unlike conventional resurgence, it is an early term/early term relation, rather than a late term/early term relation.

Here we present  the basic story of this relation. For details see  \cite{Dunne:2013ada}.  
Our goal was to understand the physical origin of the 
 resurgent trans-series in QFT  and QM.   We  decided to use elementary WKB methods to study  the origin of the exact quantization condition, which is a  generalization of the Bohr-Sommerfeld quantization condition to all orders in $\hbar$. 
The state of the art was the remarkable work of  Zinn-Justin and Jentschura (see \cite{ZinnJustin:2004ib} and refs therein). Their work proposed an exact quantization condition which involves two functions, called $A(E, g)$ and $B(E,g)$, where $E$ is the energy and $g$ the small parameter. An expansion of this (transcendental) quantization condition for small $g$ produces a trans-series expression for the energy levels. Solving  $B(E,g)=N+ \frac{1}{2}$, where $N$ is the level number for $E$, produces the usual 
Rayleigh-Schrodinger perturbation theory for the $N^{\rm th}$ unperturbed level.  $A(E,g)$   is  much more difficult to  calculate in the ZJ formalism, and it  encodes information about the instanton sector and the fluctuations around it to all orders. Despite this difficulty in calculating it, the final result of ZJ result is elegant,  and naturally incorporates all multi-instanton sectors and resurgent cancellations. We wanted to rederive it in our own way, and in the process discovered something quite surprising. 

We used  ``uniform-WKB", an approach to WKB that  is smooth across the classical turning points  \cite{Dunne:2013ada}. Since the relevant problems have harmonic minima that are very deep as $g\to 0$, one rewrites the solution in terms of  the solution of the harmonic well problem, namely the parabolic cylinder functions. Such a representation is smooth across the turning points, unlike conventional WKB. Then the global boundary condition (the Bloch boundary condition for the periodic potential, or parity for the symmetric double-well potential) combined with the known (resurgent)  asymptotics of the parabolic cylinder functions leads to the counterpart of the exact quantization condition of Zinn-Justin and Jentschura, but with a slightly different perspective.
Instead of writing   $B(E,g)$ and  $A(E, g)$,  it is more natural in the uniform WKB approach to  write $E(B, g)$ and 
 $A(B, g)$, expressing  the energy  and instanton factors as a function of the level number $B=N+ \frac{1}{2}$ and coupling $g$. This is just an innocent looking inversion, but it reveals a surprisingly simple relation between the two functions:
 \begin{eqnarray}
\frac{\partial E}{\partial B} = -\frac{g}{2S_I}  \left(2B + g \frac{\partial A}{\partial g}  \right)
\label{magic}
 \end{eqnarray}
 where $S_I$ is a numerical coefficient equal to the single-instanton action.
 This relation has a remarkable implication: given the input of $E(B, g)$,  which is equivalent to knowing perturbation theory around the perturbative vacuum, one can immediately derive, in one line, the fluctuations around any higher instanton sector. For example,   for arbitrary level number $N$,  the fluctuations around the one-instanton sector can be written to any desired order, simply given the perturbative expression to that order:
For the double-well potential the result is
\begin{eqnarray}
e^{-\frac{S_I}{g}} \textstyle{ \left[ 1 + g \frac{( -102N^2  -174 N -71)}{72} + g^2 \frac{(10404N^4 + 17496 N^3 -2112N^2 - 14172 N -6299)}{10368} + \ldots \right]}
\label{magic2}
\end{eqnarray}
and a similar result applies for the periodic potential, and others \cite{Dunne:2013ada}. Furthermore, one can also write similar relations for higher instanton sectors. The conclusion is that the two functions, B and A, which enter the exact quantization condition are not actually independent of one another: the non-perturbative function A is encoded in the perturbative function B.

This is astonishing. A direct Feynman-diagrammatic computation of the fluctuation in (\ref{magic2}), even for the $N=0$ ground state, involves many Feynman diagrams, in which the propagators are not free propagators, but propagators in an instanton background. A recent 3-loop computation \cite{Escobar-Ruiz:2015nsa}, for both the double-well and periodic potentials, has confirmed the result (\ref{magic2}) for the $N=0$ level where they were able to compute. In fact, not all of the more than 20 three-loop diagrams could be computed analytically, so their numerical result for the double-well potential was:
\begin{eqnarray}
 e^{-S_I/g} \left[ 1 - \frac{71}{72}g  - 0.6075424 g^2 + \ldots \right]
 \end{eqnarray}
 We see the 71/72 coefficient, and the next coefficient agrees with 6299/10368 to 7 decimal places.

 At the very least, this agreement suggests that there must be an easier way to compute the fluctuation in (\ref{magic2}), within the path integral Feynman diagrammatic method. 
 Why (\ref{magic}) works the way it does, and why  the relation between the instanton saddle and the perturbative saddle is so simple, are currently open questions. 
And can one prove (\ref{magic}) using purely functional integral methods?
It is apparent that there is an infinite set of relations between the infinite set of saddles in quantum mechanical systems, and that perturbation theory around the perturbative saddle can be used constructively to build all other saddles and fluctuations around them.  This is very likely related to a non-perturbative version of the Schwinger-Dyson equation (with {\it finite} + infinitesimal transformations, instead of only infinitesimal), and also to the fact that alien derivatives in the resurgent theory form an infinite dimensional free Lie algebra. At any rate, this is an extreme and constructive form of resurgence, that seems also to be computationally powerful. More thinking on this issue is clearly warranted.

 \section{Semi-classics revisited: the role of complex multivalued saddles}
  \label{HTA}
 An extremely interesting and deep problem in semi-classics is the following:  Can {\it complex} configurations contribute in a {\it semi-classical representation} of  a  real physical path integral  with real parameters (i.e., with a Hilbert space interpretation)? And, if so, what are the regularity requirements on these complex solutions? Can the solutions be multi-valued, or possibly even singular? 
 
 Surprisingly, until recently these questions have not received the attention we feel they deserve.
 Complex saddle solutions are not discussed in standard QFT textbook treatments of semi-classics and instantons  \cite{Coleman:1978ae,Schafer:1996wv}.
A recent paper gives a serious deliberation on some of these issues in the context of analytic continuation of Liouville theory \cite{Harlow:2011ny}. But it remains undecided, quote:  "We do not have a clear rationale for why this [inclusion of multi-valeud ``solutions"] is allowed". The physical concern of Ref.  \cite{Harlow:2011ny}  is that a multi-valued saddle would lead to a multi-valued observable, while observables must be single valued.  We will see that these issues get resolved with the use of resurgence. In our recent work \cite{Behtash:2015zha}, we have argued for the following  necessary and sufficient steps for the proper semi-classical treatment of path integrals of quantum mechanics:
\begin{itemize}
\item[\bf 1)]{\bf Complexification:} The action functional should be turned into a holomorphic functional of the field variables, even for real values of the coupling.  Given a Euclidean Lagrangian $\frac{1}{2} \dot x^2  
+ V(x)$, turn it into $\frac{1}{2} \dot z^2  
+ V(z)$, where $z(t)=x(t)+i y(t)$.
\item[\bf  2)]{\bf Holomorphic Newton's equation:} The saddle fields should be found by solving the holomorphic version of Newton's equation in the inverted holomorphic potential:
\begin{eqnarray}\label{eq:eom1}
\frac{d^2 z}{dt^2}=\frac{\partial V}{\partial z}  
      \qquad {\rm or \; equivalently}  \qquad   
 \frac{d^2 x}{dt^2}  = + \frac{\partial { V_{\rm r}} }{\partial x} 
\quad, \quad
 \frac{d^2 y}{dt^2}  = - \frac{\partial { V_{\rm r}} }{\partial y}   \, ,
\end{eqnarray}
where  $V(z) = V_{\rm r}(x,y) +i V_{\rm i}(x,y)$, and  the potential satisfies the 
 Cauchy-Riemann equations,  $\partial_x V_{\rm r} = \partial_y  V_{\rm i}$ and $\partial_y V_{\rm r} = -\partial_x V_{\rm i}$.
The holomorphic (Newton's) equations are of fundamental importance for semi-classics, and they are not same as the usual Newton's equation in the inverted potential: $   \frac{d^2 x}{dt^2}  = + \frac{\partial { V} }{\partial x} $. 
The generic solutions here will be complex and possibly multi-valued.  
The integration cycles attached to critical points  are found by using Picard-Lefschetz theory (the complex version of Morse theory).  
\item[\bf  3)] {\bf Perturbation theory:} Develop perturbation theory around the perturbative vacuum. This can be done by the Bender-Wu method \cite{Bender:1969si} to all orders, and generically results in asymptotic divergent series.

\item [\bf  4)] {\bf Resurgence:} The multi-valuedness of the saddles cancels against multi-valuedness of the Borel resummation of the divergent series, yielding single valued observables. We refer to this generalized summability as  Borel-Ecalle (BE)-summability. 
\end{itemize}

These four items are integral. One cannot consider any one of them in isolation and hope to get a consistent treatment of the path integral or perturbation theory. Each one of them will yield in a certain way pathological results and only in combination will one get a meaningful result. 

\subsection{Quantum mechanics of a particle with internal spin}  
 \begin{figure}
     \includegraphics[width=1.\textwidth]{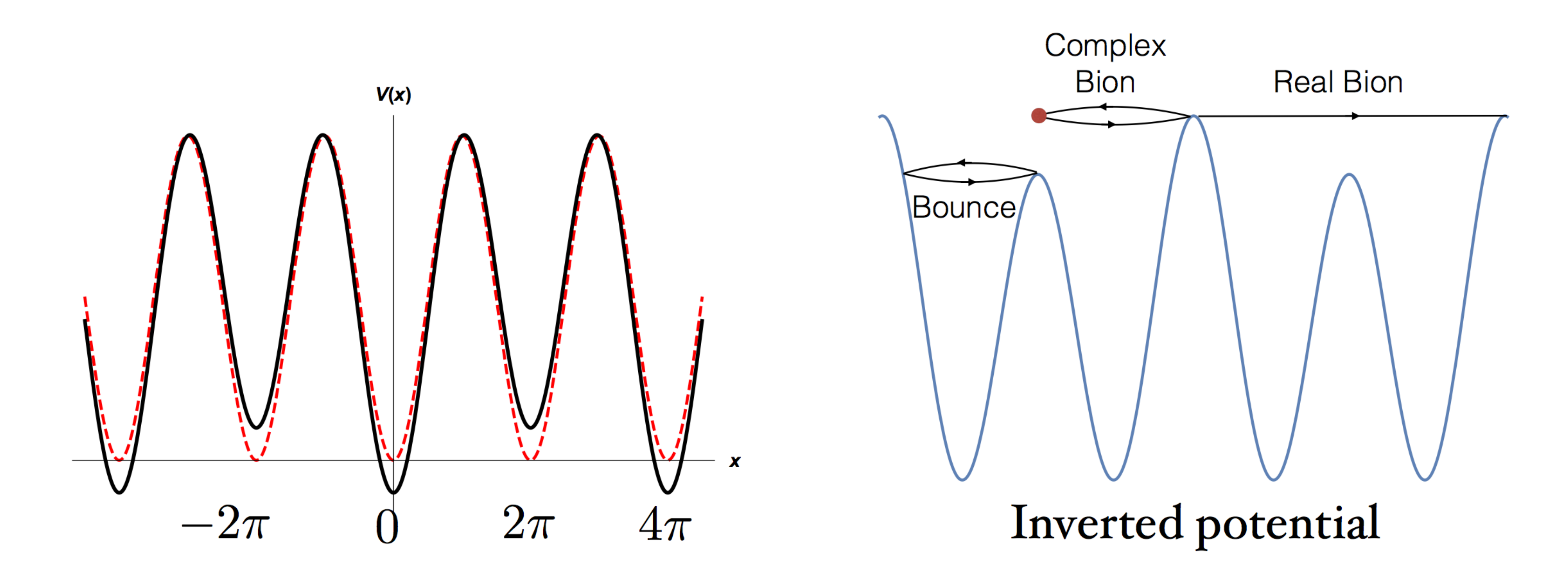}
     \caption{(Left:) Red is bosonic potential, and black is the graded potential (obtained upon projecting to a fermion number eigenstate.)   (Left:)  Inverted graded potential  and types of interesting saddles.  }
     \label{fig2}
     \end{figure} 

The clearest and most dramatic example showing the necessity of complex saddle solutions arises in the study of supersymmetric quantum mechanics. Consider a particle with position $x(t)$ and internal spin $\frac{1}{2}$, with Euclidean Lagrangian:
\begin{equation}
{\cal L}_E =  \half\dot x^2 + \half (W')^2 
            +  (\bar\psi \dot\psi   + W'' \bar \psi \psi ) 
\label{lag}
\end{equation}
Bosonize this quantum mechanics by quantizing, and then projecting fermions to spin eigenstates. This results in a sequence of graded Euclidean Lagrangians ${\cal L}_{\pm}  = \frac{1}{g} \left( \half\dot x^2 + \half (W')^2  \pm  \half g  W'' \right) $.   If we choose $W(x)$ 
to be a periodic function, for example $W(x)= \cos \frac{x}{2}$, we may identify 
$x= x+ 4\pi$ as the same physical point, corresponding to a 
 particle on a circle, rather than in an infinite lattice.

 \begin{figure}
     \includegraphics[width=1.\textwidth]{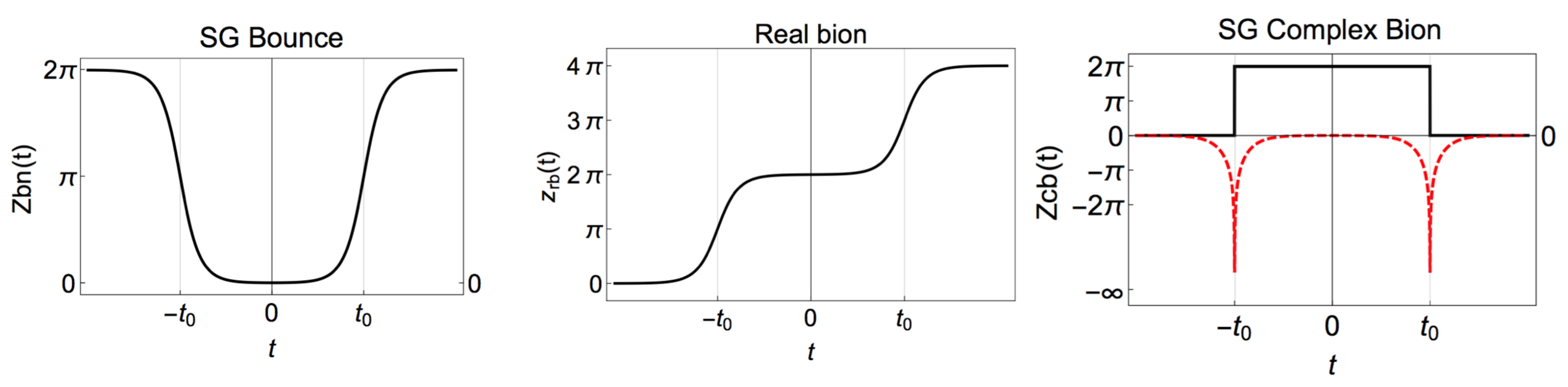}
     \caption{Exact solutions: Bounce,  real bion and complex  bion.Compex bion is multi-valued,and singular.  The other solution is complex conjugate of the one in figure.}
     \label{fig3}
     \end{figure} 
 
 This theory has  supersymmetric Witten index   $I_W=0$, yet supersymmetry is  unbroken  \cite{Witten:1982df}. It has  two supersymmetric ground states, one  in the bosonic and the other in the fermionic sector (spin up and down sectors). Thus, the ground state energy is zero to all orders in perturbation theory and also non-perturbatively: $ E_0=0$.

Let us try to reproduce this result from the completely bosonized picture without any recourse to supersymmetry.   Looking at the inverted potential, shown in Figure \ref{fig2}, there is a clear real saddle solution,  that we refer to as a real bion, interpolating from  $ [0 \rightarrow 4 \pi]$. The amplitude of the real bion, up to prefactors is $ [{\frak {RB}}] \sim e^{-S_{\rm rb}}$.   There is also a real bounce solution, but it is not related to ground state properties.  The contribution of the real bion to the ground state energy is negative. This means that if the Euclidean path integral is restricted to only real paths in a semi-classical treatment, then the supersymetry algebra  (which demands positive semi-definiteness of the energy)  and semi-classics (based on real saddles) are in contradiction.  

The resolution is that there is nothing wrong with the supersymmetry algebra, or with  
semi-classics. But we must expand our view of semi-classical analysis to include complex saddles, 
just as we do for ordinary contour integrals.
The standard semi-classical analysis, restricted to real saddles, is incomplete and will lead to inconsistencies and wrong results,  as above. 
Correct semi-classics necessitates complexification of the path integral. Once this is done, there is  a complex solution, that we call the complex bion, and it restores the consistency of supersymmetry and the semi-classical expansion \cite{Behtash:2015zha}.

For the case of  the periodic bosonic superpotential $W(x)= \cos \frac{x}{2}$, the complex bion can be found exactly  \cite{Behtash:2015zha}, and its form seems to defy almost all  standard intuition about semi-classics. The complex bion is complex, multi-valued, and  singular, and yet it has finite action and contributes to path integral.  
The real and complex bion amplitudes are 
\begin{eqnarray}
 [{\frak {RB}}] \sim e^{-S_{\rm rb}}, \qquad 
  [{\frak {CB}}] =  [{\frak {RB}}] e^{\pm i \pi}  \sim   e^{-S_{\rm rb} \pm i \pi}  
\end{eqnarray}
Note that the imaginary part of the complex bion action is $\pm \pi$.  
Thus, the contribution of the complex bion saddle to the ground state energy is actually {\emph {positive}}, the opposite sign of the contribution of the real bion saddle. (This is contrary to the folklore  which states that non-perturbative effects in bosonic systems must always reduce the ground state energy.)
Hence, the ground state energy to this leading order of semi-classics  is 
 \begin{eqnarray}
& \Delta E_0^{\rm np}   = 0 +  ( -2 -2  e^{ i \pi }  ) e^{-S_{\rm rb}}  =0
\end{eqnarray}  
Here 0 is the perturbative saddle contribution, and the negative and positive non-perturbative contributions are due to the real and complex bions, respectively. 
Complexification of the semi-classical representation of the path integral prevents a contradiction between the supersymmetry algebra and semi-classics. 

A particular deformation of the supersymmetric theory is very  enlightening \cite{Balitsky:1985in,  
Behtash:2015zha}. Deform the Yukawa term in (\ref{lag}) into 
$p W'' \bar \psi \psi$,   where $ p$ is the supersymmetry breaking deformation parameter. Then, for $p\neq 1$ the ground state energy is neither zero perturbatively nor non-perturbatively. The complex bion amplitude becomes ambiguous  $  [{\frak {CB}}] =  [{\frak {RB}}] e^{\pm i p  \pi}$
and the perturbative contribution to the ground state energy becomes non-vanishing and non-Borel summable. By resurgence, these two pathologies cancel each other, similar to 
(\ref{cancel}), but now as 
${\rm Im} {\mathbb B}E_{0,\pm} +  {\rm Im} [{\frak {CB}} ]_{\pm}   + \ldots   =0 $
Consequently, the ground state energy takes an ambiguity free form: 
 \begin{eqnarray}
& \Delta E_0^{\rm np}   = {\rm Re} {\mathbb B}E_{0}   +    [{\frak {RB}} ] + {\rm Re} [{\frak {CB}} ]  
= {\rm Re} {\mathbb B}E_{0}  +  ( -2 -2  \cos (p \pi)  ) e^{-S_{\rm rb}}  + O(e^{-2 S_{\rm rb}})
\label{interference}
\end{eqnarray}

\subsection{Hidden topological angles (HTA) and sign problem}
It is now appropriate to comment on the hidden topological angles and their importance in 
semi-classical analysis, as well as in Lefschetz thimble decompositions, both in continuum and lattice field theory and simulations. 
These angles may lead to crucial destructive/constructive interference effects  (in a Euclidean sense) among Euclidean saddles, as in (\ref{interference}). The above  examples also show that in an attempt to perform lattice simulations 
using Lefschetz thimbles, e.g.,  
\cite{Cristoforetti:2012su,Cristoforetti:2013wha,Fujii:2013sra,Aarts:2014nxa},  
all thimbles with {\it non-vanishing}  Stokes multipliers {\it must} be carefully summed over   
to correctly capture the dynamics of the theory.    Not doing so will result in erroneous results in general, especially in the presence of equally dominant saddles with different phases.  
This perspective is especially emphasized, 
with evidence from concrete examples, in \cite{Behtash:2015kna}.
It is important to note that  this is not an exponentially small  sub-leading   resurgent cancellation. 
 Rather, it is a competition between equally dominant effects in observables, as in (\ref{interference}).

For example, in the supersymmetric QM example above, if one only takes into account the real bion saddle, one 
deduces a negative ground state energy, which contradicts the SUSY algebra. If one only takes into account the complex bion saddle, one deduces positive energy, in contradiction to the fact that SUSY is unbroken in this model.  
If one just takes  into account the perturbative saddle, one will get zero, but this is an accident. For example, in a model which breaks supersymmetry,  the perturbative saddle contribution would still be zero, but the non-perturbative complex bion contribution is positive definite, and it is necessary to sum up over the two in order  to obtain the correct non-perturbative behavior \cite{Behtash:2015zha}. As emphasized, in general, one needs to add up all  saddle contributions whose Stokes multipliers are non-vanishing. 
Indeed, recent lattice and analytical  studies confirm the correctness of this perspective, also in theories with the  sign problem \cite{Nishimura:2014rxa,Kanazawa:2014qma,Tanizaki:2015rda,Alexandru:2015xva}.  Ref.~\cite{Hayata:2015lzj}  proposes a method to introduce the HTAs to fix certain pathologies of the complex Langevin method.

{\vspace{.5cm}\noindent\bf Hidden topological angle in  4d QFT:} 
${\cal N}=1$ SYM is a special theory in 4d. On one hand, it is a vector-like (QCD-like) theory (in its technical sense, a theory without elementary scalars), and it is also an integral part of SQCD. 
The gluon condensate  $\langle {\textstyle \frac{1}{N}} {\rm tr} F^2_{\mu \nu} \rangle $  is  an order parameter in this theory, and as such,  must vanish since supersymmetry is unbroken \cite{Witten:1982df}. However, vanishing gluon condensate is puzzling if one does not use tools of supersymmetry.  In Euclidean space,  integrating out fermions results in a positive definite measure, and $ \frac{1}{N} {\rm tr} F^2_{\mu \nu}$ is also positive definite.  Then, how can one explain 
the vanishing of the condensate from  a  semi-classical point of view?  

\begin{figure}[ht]
\begin{center}
\includegraphics[angle=0, width=\textwidth]{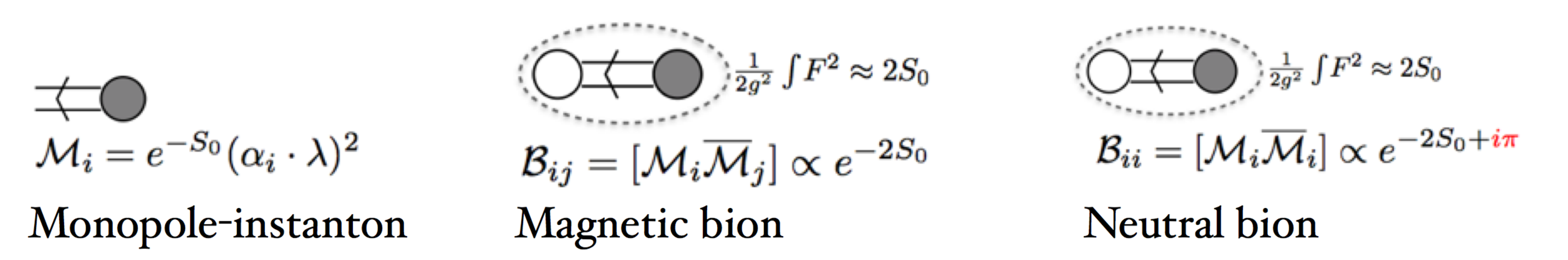}
\caption{Saddles in   ${\cal N}=1$ SYM. Note the  HTA associated with  neutral bion. }
\label {Thimbles-SYM}
\end{center}
\end{figure} 
The resolution is relatively counter-intuitive.  Although this is a problem {\it without} a sign problem in the space of real fields,   the path integral needs to be holomorphized as in the QM example, and as such, it will now possess a sign problem!  Then, in semi-classics, there are saddles with complex phases.    In order to do semi-classics reliably, compactify the theory on a 
small  $\mathbb R^3 \times S^1_L$, where the theory becomes weakly coupled, and yet 
continuously connected to $\mathbb R^4$ (for periodic boundary conditions for fermions).  
It has three types of interesting saddles. These are:  
 monopole-instantons, with amplitude ${\cal M}_i =  e^{-S_0}  (\alpha_i \cdot \lambda)^2 $,   magnetic bions  with amplitude  ${\frak B}_{ij}= [{\cal M}_i \overline {\cal M}_j]
      = e^{-2S_0}\ldots  $,   and   neutral bions,  ${\frak B}_{ii}= [{\cal M}_i \overline {\cal M}_i]
      = e^{-2S_0 +i \pi}\ldots$.   
where $\alpha_i,\;i=1, \ldots, N$  is an element of the  affine-root system,  and ${\frak B}_{ij}$ and ${\frak B}_{ii}$ are 
non-vanishing $\forall \hat A_{ij} <0$, and  $\forall \hat A_{ii} >0$ 
entries of the extended Cartan matrix.  
Note that the monopole action is $S_0= \frac{8 \pi^2}{g^2N}=\frac{S_I}{N}$, thus these saddles are exponentially more important than the 4d instanton with action $S_I$. Since monopoles have fermion zero modes, they do not contribute to  $\langle {\textstyle \frac{1}{N}} {\rm tr} F^2_{\mu \nu} \rangle $.  
But magnetic and neutral bions do, and their densities are the same! However, 
 there  is a  $\mathbb Z_2$ worth of relative phase between the two saddles/thimbles: 
\begin{eqnarray} 
{\rm Arg} ( \mathcal{ J}_{ {\frak B}_{ii}} )  
    = {\rm Arg} (\mathcal{J}_{ {\frak  B}_{ij}} )  +  \pi\, 
\end{eqnarray}
Consequently, contrary to the folklore  which states that non-perturbative contributions to the gluon condensate must be positive, there exists both positive and negative contributions: 
\begin{eqnarray}
\label{cancel2}
 \langle {\textstyle \frac{1}{N}} {\rm tr} F^2_{\mu \nu} \rangle  
  \propto  
    (n_{ {\frak  B}_{ij}} +   e^{i  \pi}n_{ {\frak B}_{ii}})  = e^{-2S_0} +  e^{-2S_0+i \pi}  =0 \, . 
\end{eqnarray}
This  reflects the importance of summing over all active saddles in the Lefschetz thimble decomposition (\ref{complex-2}).   Otherwise semi-classical analysis would result in  a contradiction with the supersymmetry algebra; see \cite{Behtash:2015kna, Behtash:2015zha} for details, and also non-supersymmetric examples.

\section{QFT:  Sigma models,  gauge theories and renormalon puzzle}
One may hope  that the mechanism of resurgent cancellation  that applies so nicely in QM also applies to asymptotically free QFT,  such as the two-dimensional ${\mathbb {CP}}^{N-1}$ sigma model, but it is well-known that this is not the case. More than three decades ago, 't Hooft argued that the
Borel transform of perturbation theory  not only has Borel-plane singularities at  multiples of $t= 2S_I$, but also far more dominant singularities at $t^{\rm IR}_{n}= n  \frac{ 2S_I}{\beta_0}$, on the positive real axis 
$ \mathbb R^{+}$  in the Borel plane, called IR-renormalon singularities.  

Of course, the usual semi-classical method does not apply on $\mathbb R^2$ for sigma models. In particular, the so called ``dilute instanton gas approximation" is not a well-defined approximation on $\mathbb R^2$. The instantons have a size modulus, and the first assumption of the dilute instanton gas, that the separation between instantons must be much larger than the instanton size, is violated.  

The instanton size modulus problem can be controlled by compactifying the theory on a small circle, but doing so in a conventional (thermal) way results in a thermal regime of the sigma model  in which the microscopic degrees of freedom are ungapped and liberated. In fact, the free energy scales as $NT^2$, 
the usual Stefan-Boltzman law, and such a small-circle regime is not sufficiently related to the large-circle or infinite volume regime.  Instead,  we use an idea that has first developed in the context of gauge theory \cite{Unsal:2008ch}.
It is the idea of adiabatic continuity. In gauge theory on $\mathbb R^3 \times  S^1$,  the small circle theory can be deformed by a double-trace operator such that the small-circle and large-circle physics are continuously connected. This can also be achieved by having adjoint fermions and endowing them with periodic (rather than anti-periodic) non-thermal boundary conditions 
\cite{Unsal:2007jx,Bergner:2014dua}. 
The counter-part of this step in sigma models is the   {\it twisted boundary conditions} , where fields are not periodic on $S^1$, rather  rotated by  a unitary  twist matrix 
$\Omega= {\rm Diag}  ( 1, e^{ i \frac{2 \pi}{N}} ,  \ldots, e^{ i \frac{2 \pi(N-1)}{N}})$,
 which guarantees that the small and large-circle regimes are adiabatically connected.

 \begin{figure}
     \includegraphics[width=1.\textwidth]{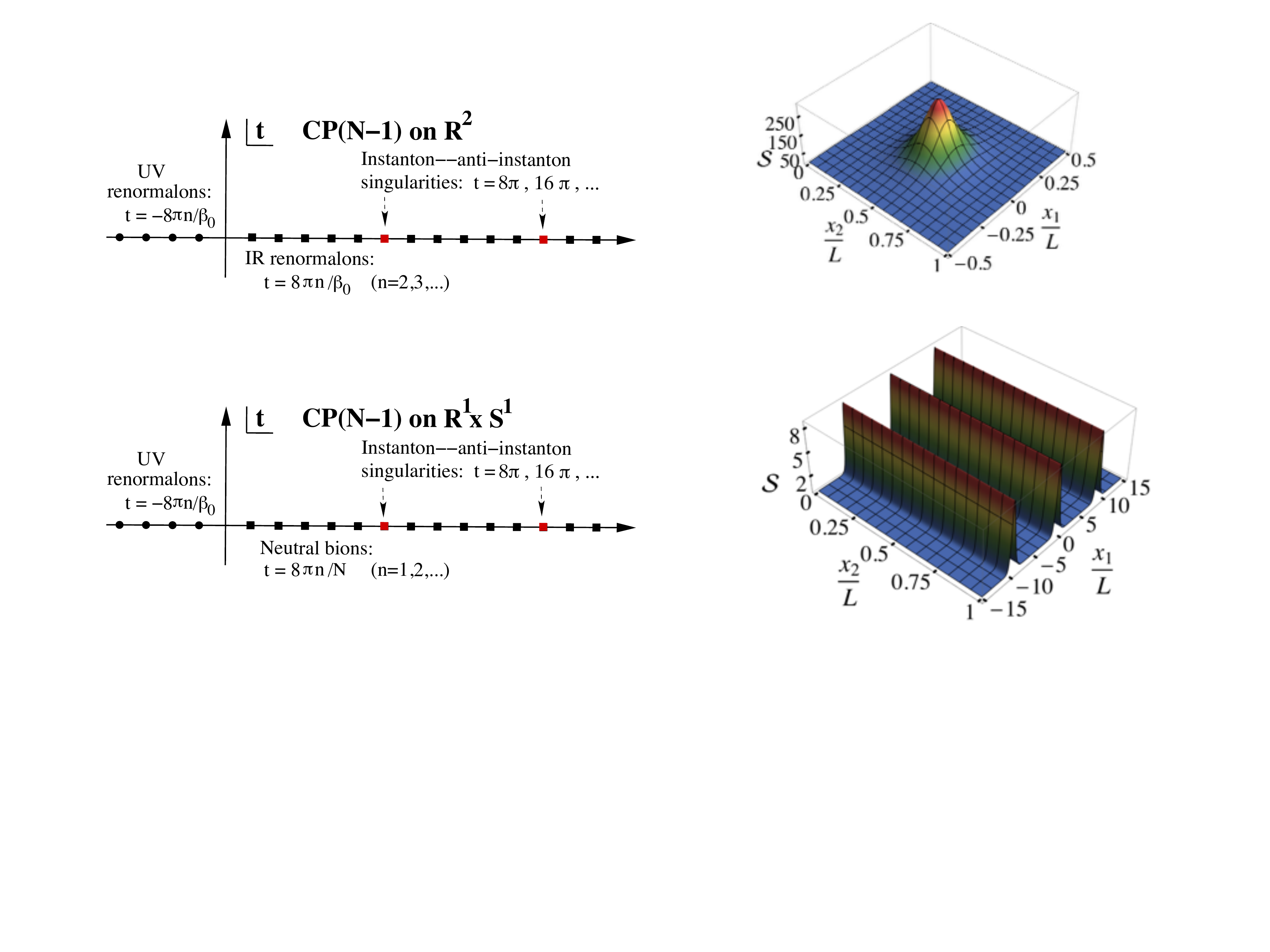}
  \vspace{-4cm} 
      \caption{(Left) Borel plane for the ${\mathbb {CP}}^{N-1}$ model on $\mathbb R^2$  and  $\mathbb R^1 \times S^1$. (Right) Fractionalization of instanton into  $N$ fractional instantons upon  imposing twisted boundary conditions. Figure is for ${\mathbb {CP}}^{N-1}$ with $N=4$.}
     \label{fig4}
     \end{figure}

Once these twisted boundary conditions (tbc)  are used,  several remarkable  things happen \cite{Dunne:2012zk,Misumi:2014jua}.  It is useful to represent what happens in contrast to periodic boundary conditions (pbc). 
\begin{itemize}
\item With tbc, the quantization of momentum in the compact direction becomes in units of $\frac{2 \pi}{LN} $ rather than the usual Kaluza-Klein momentum  $\frac{2 \pi}{L} $, which is the case with pbc. The $\frac{2 \pi}{LN} $ spacing is very peculiar, because it tells you that even at finite-$L$, you can turn the  perturbative spectrum into the continuum one, exactly as in  $\mathbb R^2$. This is indeed an imprint that at large-N, these compactifications satisfy large-N volume independence (Eguchi-Kawai  or large-N reduction).  This fact already tells you that the small-circle limit obtained in this manner must be special.

\item With pbc, the small-circle limit is described by a quantum mechanical particle on the ${\mathbb {CP}}^{N-1}$ manifold,  e.g. particle on $S^3$ for $N=2$. With tbc, the QM is a particle on ${\mathbb {CP}}^{N-1}$, as well as  a potential on it with $N$ minima,  e.g. particle on $S^3$ for $N=2$, where the north and south pole are minima of a potential, with a  barrier in between.  This additional potential has an interesting implication for instantons.

\item With pbc, the instanton of the theory on   $\mathbb R^2$ does not fractionalize.  With tbc, the instanton of the theory  fractionalizes into $N$ kink-instantons.  The $i^{\rm th}$ kink-instanton action is $S_{\rm i}= \frac{S_I}{N}= \frac{S_I}{\beta_0}$.   The size modulus problem is fixed by tbc in a good way, see below.  The kink-instantons ${ \cal K}_i, i=1, \ldots N$ are the leading saddles and are associated with the affine-root system of the $\frak {su}(N)$ algebra.  

\item The mass gap in the reduced QM is a kink-instanton effect and is given by  $m_g \sim (LN)^{-1} e^{-S_{\rm i}} =  (LN)^{-1}  e^{- \frac{S_I}{\beta_0}
} = \Lambda$, where  $\Lambda$ is the strong scale of the theory. This means, the non-perturbative small-circle dynamics remembers the strong scale of the theory, and in fact, this is nothing but the (expected) mass gap in the decompactification limit.  In contrast, the long distance dynamics in pbc totally forgets  the strong scale.   The "unforgetfulness"  of  tbc  is  remarkable.

\item  
Now comes the issue about renormalons. Of course, IR-renormalons on $\mathbb R^2$ are  due to phase space integration, from the exponentially low momentum region.  In reduced QM (with tbc), the renormalon type divergence  $\sim \frac{n!}{(2S_I)^n}$ is reproduced by the   combinatorics 
\cite{Anber:2014sda,Dunne:2012zk}.  
 (By contrast, pbc with regular reduction has no memory  of the IR-renormalon singularity.) Studying   the large-orders of perturbation theory for the reduced theory, and  its Borel resummation, one finds that the ambiguity in the Borel resummed perturbation theory is of order $e^{-  \frac{2S_I}{\beta_0}}$,  with a singularity at Borel plane  at $t^{\rm IR}=  \frac{ 2S_I}{\beta_0}$.  

\item
 At second order in the semi-classical expansion, the saddle structure is universal and dictated by  $\frak {su}(N)$ algebra data. For all non-vanishing entries of the extended Cartan matrix $\hat A_{ij}$,  there exists a  non-selfdual saddle,  charged  bions   $[\frak B_{ij}] =[{ \cal K}_i \bar { \cal K}_j] $ for 
$\hat A_{ij} <0$, and  neutral bions,  
$[\frak B_{ii}]_{\pm}  =[{ \cal K}_i \bar { \cal K}_i]_{\pm} $ for all $\hat A_{ii} >0$.  Neutral bions, as in  our QM analysis, are two-fold ambiguous.

\item
The imaginary ambiguity in neutral bions  ($\sim \pm i e^{-  \frac{2S_I}{\beta_0}}$)  cancels exactly the ambiguity in  Borel resummed perturbation theory, ${\rm Im} {\mathbb B}E_{0,\pm} + [\frak B_{ii}]_{\pm} = 0$.   The location of the neutral bion ambiguity at small 
$\mathbb R^1 \times S^1$ is exactly the location of the elusive 't Hooft IR-renormalon ambiguity 
for the theory defined on $\mathbb R^2$. This is the sense in which our neutral bions are continuously connected to the 
IR-renormalons (or what becomes of IR-renormalons in the small-circle weak coupling regime).  
 See Fig. \ref{fig4}, left.

\end{itemize}

{\bf Non-selfdual saddles and sigma models without instantons:} Perhaps even more interesting than this resurgent cancellation in the 2d ${\mathbb {CP}}^{N-1}$ sigma model, between ambiguities of Borel non-summable perturbation theory and certain non-perturbative neutral bion objects, is the fact that this also happens in other sigma models which have no instantons. For example, the Principal Chiral Model  and the $O(N)$ sigma model have the same IR renormalon problems in the perturbative sector, but have no instantons. At first sight this would seem to doom to failure the resurgent cancellation mechanism. However, these sigma models have finite action saddle solutions, solutions to the second order Euclidean equations of motion, and their action is quantized. These classical solutions had been neglected previously as being unphysical, as they have negative fluctuation modes, but  in fact they play a role similar to that of an instanton/anti-instanton configuration in the ${\mathbb {CP}}^{N-1}$ model.
With tbc in the compactified theory these non-BPS saddles fractionalize, in just the correct manner to correspond to the beta function coefficient.  These fractionalized saddles are more dominant in the semi-classical expansion, and lead  to the cancellation of ambiguities arising from the Borel non-summability of the perturbative sector \cite{Cherman:2013yfa}.
 \\
 


{\bf Acknowledgments:}
We acknowledge support from DOE grants DE-SC0013036 (M.\"U.) and DE-SC0010339 (G.D.).
M.\"U.'s work was partially supported by the Center for Mathematical Sciences and Applications  at Harvard University.

\end{document}